# Link Prediction by De-anonymization: How We Won the Kaggle Social Network Challenge

Arvind Narayanan     Elaine Shi     Benjamin I. P. Rubinstein

*Abstract*—This paper describes the winning entry to the IJCNN 2011 Social Network Challenge run by Kaggle.com. The goal of the contest was to promote research on real-world link prediction, and the dataset was a graph obtained by crawling the popular Flickr social photo sharing website, with user identities scrubbed. By de-anonymizing much of the competition test set using our own Flickr crawl, we were able to effectively game the competition. Our attack represents a new application of de-anonymization to gaming machine learning contests, suggesting changes in how future competitions should be run.

We introduce a new simulated annealing-based weighted graph matching algorithm for the seeding step of de-anonymization. We also show how to combine de-anonymization with link prediction—the latter is required to achieve good performance on the portion of the test set not de-anonymized—for example by training the predictor on the de-anonymized portion of the test set, and combining probabilistic predictions from de-anonymization and link prediction.

## I. INTRODUCTION

KAGGLE.COM—a platform for machine learning competitions—ran the IJCNN 2011 Social Network Challenge for 9 weeks from Nov 8, 2010 through Jan 11, 2011 [18]. The goal of the Social Network Challenge was to promote research on link prediction. The contest dataset was created by crawling a large online social network and partitioning the obtained edge set into a large training set and a smaller test set of edges augmented with an equal number of fake edges. Challenge entries were required to be probabilistic predictions on the test edge set. Node identities in the released data were obfuscated to prevent cheating.

Of the 119 teams competing in the link prediction task, we placed first by de-anonymizing a large portion of the test set and applying a combination of standard machine learning techniques on the remainder. While our participation in the challenge raised obvious questions on the propriety of our methods, our goal was to raise attention to the ever-present possibility of de-anonymization in such contests.[1]

The contributions of this paper are three-fold. First we demonstrate that partial crawls of a large real-world online social network can be effectively de-anonymized, whereas prior work studied de-anonymizing complete snapshots of social networks [26]. We achieve this by focusing on nodes with high in-degrees for "seeding" the de-anonymization process. As we explain in Section III-A, the set of high in-degree nodes is (approximately) preserved even in a snapshot obtained from a partial crawl.

Second, we formulate seed identification—the first step of de-anonymization—as a combinatorial optimization problem, specifically *weighted graph matching*, in contrast to the pattern search approaches of [6] and [26]. We then show how to use simulated annealing to solve this problem. Since our formulation makes no assumptions specific to the de-anonymization context, our solution is broadly applicable to the weighted graph matching problem.

Third, our winning entry, which yielded a combined test Area Under Curve (AUC) of 0.981, made use of a novel combination of standard link prediction with de-anonymization to game a popular link prediction contest. Moreover the link prediction component of our entry was advantaged by training on the de-anonymized portions of the test set. While previous applications of de-anonymization have been to privacy attacks [27], [25], to the best of our knowledge this is the first application of de-anonymization to gaming a machine learning contest.

The success of our approach has important consequences for future machine learning contests particularly in social network analysis. We argue that while appropriate contest rules should be used to disincentivize gaming through de-anonymization, technical measures to detect gaming are unlikely to be foolproof. Releasing useful social network graph data that is resilient to de-anonymization remains an open question.

## II. BACKGROUND

In this section we describe the competition dataset and our Flickr dataset crawled for the purposes of de-anonymization.

### A. Social network challenge dataset

The challenge graph had 1,133,547 nodes and 7,237,983 edges. An additional 8,960 edges formed the test set; of these, equal numbers were true edges (withheld from the training set) and false edges. Test edges were incident on training set nodes only, making link prediction possible. Of the 1.1m nodes, only 37,689 had outgoing edges. In the sequel, we will call these "crawled" nodes. In order to generate a running leaderboard, a random 20% of the test set was held out to form a validation set on which entry AUCs were computed. The final results were calculated on the entire test set.

A. Narayanan is with the Department of Computer Science, Stanford University (email: arvindn@cs.utexas.edu).
E. Shi is with the Computer Science Division, UC Berkeley and PARC (email: elaines@cs.berkeley.edu).
B. Rubinstein is with Microsoft Research, Silicon Valley (email: ben.rubinstein@microsoft.com).

[1]Once we had established a clear lead on the public leaderboard, we contacted the organizers to inquire about how to proceed, offering to withdraw if that was deemed to be best. Happily, the organizers found our method acceptable, and noted its novelty.

User identities were obfuscated in the challenge dataset: nodes were assigned random numeric IDs. However, during the course of the challenge, it was revealed on the contest forum that the data came from crawling Flickr—nodes correspond to users of the photo sharing site and edges correspond to directed friendship relationships. After completion, we learned that the competition crawl took place between 21–28 June 2010 [24]. It was also revealed that the crawl was made in several iterations by initially downloading random nodes, and subsequently sampling uniformly from the inbound neighbors of these nodes; that the true test edges were selected uniformly at random from the crawled set of edges incident to nodes with degree two or more; and that the false test edges were rejection sampled to connect a random crawled node to a random reached node, provided the edge did not already exist.

Figures 1 and 2 display the in- and out-degree distributions of the competition training graph, respectively. The heavy tailed distributions suggest that a small number of nodes have very large degrees. These nodes form the seed set for the de-anonymization process below.

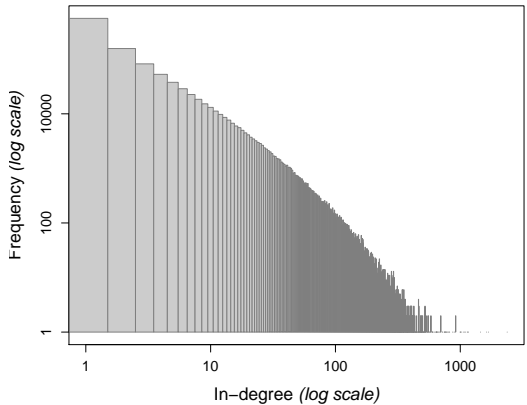

Fig. 1

KAGGLE CHALLENGE GRAPH: IN-DEGREE DISTRIBUTION.

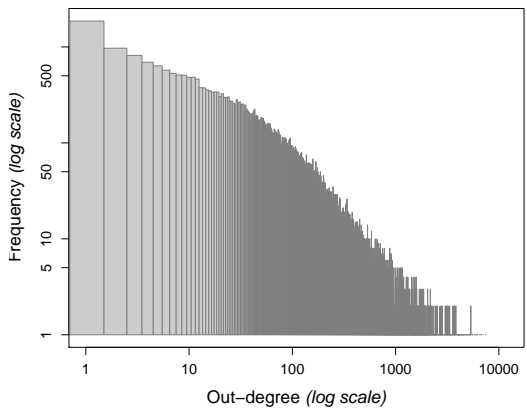

Fig. 2

KAGGLE CHALLENGE GRAPH: OUT-DEGREE DISTRIBUTION.

*B. Flickr crawl dataset*

Between mid-December 2010 and mid-January 2011, we crawled the Flickr social graph in order to de-anonymize the Kaggle dataset. The crawler was written in Python, using the Curl library. A total of 2,038,803 nodes were crawled; these are the nodes that have outgoing edges. The resulting graph comprises 163,579,517 directed edges on 9,124,801 nodes—significantly larger than the competition dataset. Of these 9.1m nodes, 7,775,972 have incoming edges.

While unhelpful for de-anonymization and link prediction, we recorded the true identities of the nodes we crawled. Similarly the organizers provided us with the unobfuscated node IDs for the challenge graph after the contest. Together these identities allowed us to compare the two partial snapshots of the Flickr graph after the completion of the contest.

It is only meaningful to compare edges originating from nodes that have been crawled in both graphs. There were 6,658,755 such edges in the challenge graph, 7,041,554 edges in our Flickr graph and 6,545,560 of these edges were in common. Thus 113,195 edges had been deleted since the Kaggle crawl and 495,994 edges had been added. The cosine similarity between the two sets of edges is 95.6%.[2]

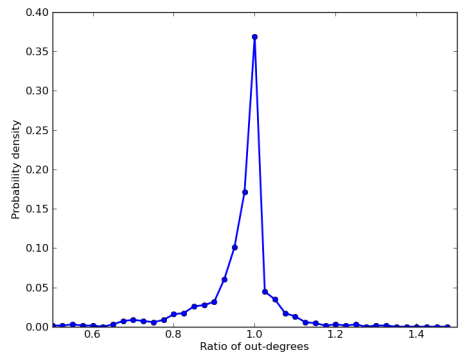

Fig. 3

DISTRIBUTION OF RATIOS OF OUT-DEGREE OF KAGGLE NODES TO CORRESPONDING FLICKR NODES. (TOP ~700 NODES BY OUT-DEGREE.)

III. DE-ANONYMIZATION

**De-anonymization of social networks** has been well studied in the security and privacy community. We summarize the literature in Section VI. Our presentation here follows Narayanan and Shmatikov [26] closely.

Abstractly, de-anonymization can be formalized as follows: there is a graph $G$ from which two graphs $G_T$ (for "target") and $G_A$ (for "auxiliary") are derived via some stochastic process. There is thus a natural notion of (partial) correspondence between the nodes of $G_T$ and $G_A$; the goal of de-anonymization is to recover this correspondence.

---

[2]Here and in the sequel, the cosine similarity between two sets $X$ and $Y$ is $\frac{|X \cap Y|}{\sqrt{|X| \cdot |Y|}}$.

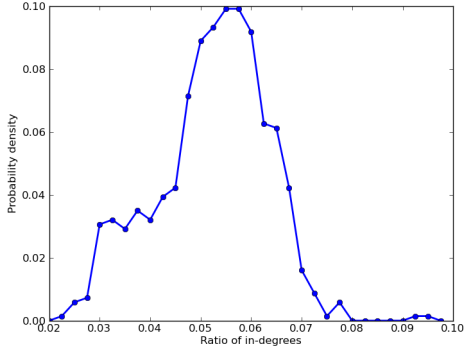

Fig. 4

DISTRIBUTION OF RATIOS OF IN-DEGREE OF KAGGLE NODES TO CORRESPONDING FLICKR NODES. (TOP ~700 NODES BY IN-DEGREE.)

The stochastic process involved could be as simple as the deletion of random edges. At the other extreme, if we're considering two entirely different online social networks, say Facebook and LinkedIn, then $G$ is the underlying social graph of human relationships, and $G_T$ and $G_A$ are generated according to the processes by which users join online social networks, which has no simple algorithmic description.

In the current instance, $G$ is the Flickr graph at the time of the Kaggle crawl. The Kaggle graph $G_T$ was generated by crawling, followed by the sampling process described in Section II. Our graph $G_A$ was "generated" from $G$ by the natural evolution of the Flickr graph since the time of the Kaggle crawl until the time of our crawl, followed by crawling. Each of these steps introduces noise and/or bias into the graphs.

**Metrics.** We use the metrics *accuracy* and *coverage* to measure the performance of a de-anonymization algorithm. Accuracy is defined as the fraction of nodes correctly de-anonymized among all de-anonymized nodes. Coverage is defined as the fraction of nodes de-anonymized. Specifically, we are concerned with coverage over the Kaggle test set. Therefore in the sequel, unless explicitly noted otherwise, coverage means the fraction of nodes de-anonymized among the nodes that appeared in the Kaggle test set. We also adopt a similar notion of *edge coverage* and *edge accuracy* for the Kaggle test set, where the former is defined as the fraction of edges in the Kaggle test set that have been de-anonymized, and the latter is defined as the fraction of edges that are correctly de-anonymized among all edges that are de-anonymized in the Kaggle test set. An edge is de-anonymized *iff* both nodes incident to the edge are de-anonymized.

**Overview of our de-anonymization algorithm.** Our basic approach to de-anonymization is described in [26]. Broadly, there are two steps: "seed identification" and "propagation." In the former step we somehow de-anonymize a small number of nodes, and in the latter step we use these as "anchors" to propagate the de-anonymization to more and more nodes. In this step the algorithm feeds on its own output.

We note some general features of this approach. There is some similarity with the "seed-and-extend" paradigm for sequence alignment in bioinformatics [5]. A very small number of seeds is sufficient to get the algorithm underway; the number depends on whether or not the seeds are "hubs" (high-degree nodes) and on the degree of similarity between the two graphs.

Propagation is somewhat reminiscent of the spread of epidemics. As this analogy suggests, if there are too few seeds, then propagation dies out; if there are sufficiently many, it reaches a large fraction of the nodes. Note, however, that while epidemics are transmitted from one node to another, multiple already-mapped pairs of nodes are involved in implicating a new mapping between a new pair of nodes.

*A. Seed identification*

As mentioned earlier, seed identification is a bootstrapping step with the goal of de-anonymizing a small subset of nodes in the Kaggle graph.

The seed-identification technique in [26] is based on pattern search, specifically, identifying small cliques based on degrees and common-neighbor counts. We propose a new approach to seed identification based on formulating it as a combinatorial optimization problem. Compared to [26], we believe our approach is more robust to graph evolution, noise, and information loss due to the crawler's sampling process.

Our motivation for developing this new approach stems primarily from the fact that during the contest, it was unclear whether the clique search technique would be feasible, since we did not know how much the graph had evolved between the two crawls, and how similar the node degrees in the two graphs were to each other. Although we determined after the contest that the existing technique would also have worked in this setting, our method may be of independent interest, especially in contexts where noise and information loss are more prominent.

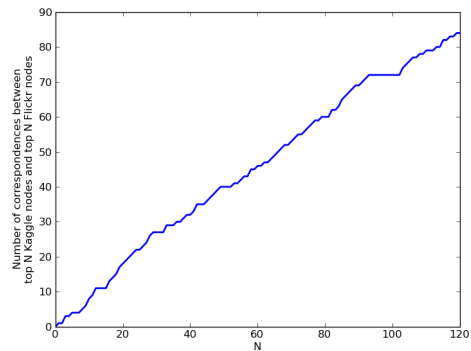

Fig. 5

CORRESPONDENCE BETWEEN KAGGLE AND FLICKR HIGH IN-DEGREE NODES.

**Search space reduction.** Since both graphs contain millions of nodes, our algorithm needs to effectively reduce the search space to be viable. Essentially, we would like to find a small subset of nodes $K$ in the Kaggle graph, and a small subset of nodes $F$ in the Flickr graph, such that a significant fraction of nodes in $K$ correspond to nodes in $F$.

The key observation is that the nodes with high *in-degree* in both graphs roughly correspond to each other (whereas only about $\frac{1}{15}$ of nodes with high out-degree are present in the Kaggle graph). Figure 5 shows the correspondence: of the top 30 nodes in the Kaggle graph and the top 30 nodes in the Flickr graph, there are 27 correspondences; 60 in the top 80 nodes and 84 in the top 120.

The fact that a rough correspondence exists can be surmised without the benefit of hindsight. If we assume that the crawled nodes in the Kaggle graph are a random sample of all nodes, ignoring the changes in the graph between the two crawls, we can expect that the indegrees of corresponding nodes in the two graphs will be roughly in the same proportion up to sampling error. Even though the Kaggle nodes are not a uniformly random sample, the sampling turns out to be "random enough" for the correspondence to hold.

**Graph matching.** We now have a small subset of nodes $K$ from the Kaggle graph, and a small subset of nodes $F$ from the Flickr graph, with a large fraction of nodes in $K$ corresponding to nodes in $F$. Our next step is to find this correspondence.

First we need a measure of the quality of a candidate mapping, so that we can optimize that measure over all possible mappings. If all the edges among the top $k$ nodes in the Kaggle graph were available, we could look for the mapping that minimizes mismatches. By mismatch we mean the existence of an edge between a pair of nodes in one graph, but not between the images of that pair of nodes in the other graph. However, since only a (small) subset of the Kaggle nodes are crawled, most of the edge information is unavailable.

It turns out that even with a partial crawl, we can adopt a similar strategy, and in fact do even better than simply looking at edges. The trick is to look at the *cosine similarity of the sets of in-neighbors* of a pair of nodes. A similar rationale as above explains why the cosines between corresponding node pairs in the two graphs would be roughly equal: if the Kaggle nodes are a random sample then they would indeed be equal up to sampling error.

Figure 6 shows the relationship between similarity scores of corresponding node pairs. Among the top 40 nodes in the Kaggle and Flickr graphs, there are 32 correspondences; these two sets of 32 nodes give $\binom{32}{2} = 496$ pairs of corresponding unordered node pairs. For each node pair we have an in-neighbor cosine similarity score; thus, the graph shows 496 pairs of scores $(x_i, y_i)$.

The slope is less than 1, i.e., node pairs in the Kaggle graph have slightly higher scores. This is because the crawled nodes in the Kaggle graph are more biased towards high-degree nodes.

We treat the cosine score of a node pair as the weight of an (undirected) "edge" between those two nodes, regardless of

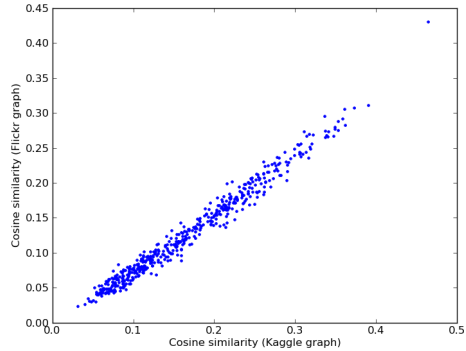

Fig. 6

RELATIONSHIP BETWEEN SIMILARITY SCORES OF CORRESPONDING NODE PAIRS.

whether or not any actual edges exist between those nodes. We look for a mapping where the weights of corresponding edges are as close to each other as possible. This is the problem of *weighted graph matching*. As mentioned earlier, this approach to finding seeds is a *global optimization problem* rather than pattern search.

During the contest, we found a seed mapping of 10 nodes among the top 20 nodes (there are in fact 18 corresponding pairs of nodes among the top 20s) by visual inspection of the matrix of cosines. This formed our seed set, and was sufficient to kick off propagation.

However, there is an automated, robust and scalable approach to the weighted graph matching problem: i.e., using a state-space search metaheuristic. In Section IV we show that simulated annealing can easily handle inputs of up size up to 100 with low false-positive and false-negative rates.

*B. Propagation*

Our propagation algorithm is adapted from [26]. In fact, our implementation was simpler; we did not need the more complex heuristics because the two graphs were substantially similar to each other, as noted in Section II. However, the fact that neither graph was fully available made the algorithm somewhat trickier.

As the algorithm progresses, it maintains a (partial) mapping between nodes in the Kaggle graph and nodes in the true Flickr graph. We iteratively try to extend the mapping as follows: pick an arbitrary as-yet-unmapped node in the Kaggle graph, find the "most similar" node in the Flickr graph, and if they are "sufficiently similar," they get mapped to each other.

At a high level, similarity between a Kaggle node and a Flickr node is defined as cosine similarity between the already-mapped neighbors of the Kaggle node and the already-mapped neighbors of the Flickr node (nodes mapped to each other are treated as identical for the purpose of cosine comparison).

In Figure 7, the blue nodes have already been mapped. The similarity between $A$ and $B$ is $\frac{2}{\sqrt{3}\cdot\sqrt{3}} = \frac{2}{3}$. Whether or

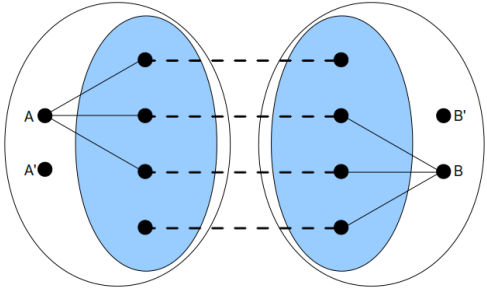

Fig. 7

DE-ANONYMIZATION PROPAGATION ALGORITHM (SIMPLIFIED).

**Input**:
- $v_K$: a node in the Kaggle graph
- $v_F$: a node in the Flickr graph
- $D$: the set of Kaggle nodes de-anonymized so far
- map: a partial map from the Kaggle to the Flickr nodes with domain $D$
- $C_K$: the set of crawled Kaggle nodes, i.e., subset of Kaggle nodes having outgoing edges
- $C_F$: the set of crawled Flickr nodes

**Output**: a similarity score

$N_K \leftarrow \mathsf{map}[\mathsf{N}^-(v_K) \cap D] \cap C_F$
$N_F \leftarrow \mathsf{N}^-(v_F) \cap \mathsf{map}[C_K]$
**if** $v_K \in C_K$ and $v_F \in C_F$ **then**
  $N_K \leftarrow N_K \cup \mathsf{map}[\mathsf{N}^+(v_K) \cap D]$
  $N_F \leftarrow N_F \cup (\mathsf{N}^+(v_F) \cap \mathsf{map}[D])$
**end**
**return** $\mathsf{CosineSim}(N_K, N_F)$

**Algorithm 1:** Similarity scores for de-anonymization. Note: $\mathsf{N}^+(v)$ and $\mathsf{N}^-(v)$ denote the out- and in-neighborhood of a node $v$ respectively.

not edges exist between $A$ and $A'$ or $B$ and $B'$ is irrelevant.

Edge directionality combined with the fact that not all nodes have been crawled introduces some subtleties. Essentially, when comparing a pair of nodes, we ignore the out-edges of either unless *both* have been crawled. The full algorithm is listed as Algorithm 1.

There are two reasons why the similarity between a node and its image may not be 100%: the contest graph is slightly different from our newer crawled graph, and the mapping itself might have inaccuracies. The latter effect is minimal—the algorithm occasionally revisits already-mapped nodes to correct errors in the light of more data.

The propagation algorithm was run in two stages. In the first stage, we de-anonymize high-degree nodes with a high confidence level. In the second stage, we focus on the nodes in the test set (that have not yet been de-anonymized) and relax the confidence threshold, and even allow multiple candidates per node. Mappings found in this stage do not feed back into the similarity computation, due to lower confidence.

The following heuristic criteria define the "sufficiently similar" criterion in stage 1 of propagation:
- There must be at least $k = 4$ pairs of common neighbors mapped to each other;
- The similarity score must be at least $\theta = 0.5$; and
- The difference in similarity scores between the best and the second-best matches must be at least $\delta = 0.2$.

The parameters $k, \theta$ and $\delta$ allow us to trade off accuracy for coverage. In the second stage of propagation, we set $k = 3$, eliminate the last criterion, and if there are multiple matches with a similarity score of $0.5$ or more, we report the best 3 as candidates.

*C. Results*

The algorithm was implemented in Python and run on a commodity laptop. Propagation was run in two stages. In the first stage, we attempted to de-anonymize the high-degree nodes, specifically, about 34,000 nodes sorted by out-degree and about 100,000 nodes sorted by in-degree. There was, of course, some overlap between the two sets. Mappings were produced for about 120,000 nodes. In the second stage, we attempted to de-anonymize the nodes in the test set that hadn't already been mapped. The first stage took a few hours and the second stage about 20 minutes.

Of the 120,000 mappings in stage 1, 99.3% were correct.[3] At the end of the second stage, mappings were produced for about 14,000 nodes out of about 17,600 in the test set. Thus, the coverage was 79.7%. About 7.5% of nodes had multiple candidates. Of these, the top match was accurate in the overwhelming majority of cases. Considering only the top match, the overall accuracy was 97.8%. Of the 8,960 edges, the coverage was 64.7% (roughly the node coverage squared), and the accuracy was 95.2%.

IV. GRAPH MATCHING VIA SIMULATED ANNEALING

In Section III-A we showed how to formulate the problem of seed-identification, the first step of de-anonymization, as a *weighted graph matching* problem. To recap, we are given two graphs with weighted edges, and we are required to find the mapping between the nodes that minimizes a given function, such as the sum of absolute differences between the weights of edges mapped to each other.

Weighted graph matching generalizes its unweighted version, *inexact graph matching*, which in turn generalizes graph isomorphism. Unsurprisingly, inexact graph matching is NP-complete [2], and therefore weighted graph matching is NP-complete as well.

**Choice of weights.** In our experiments the weight of an edge is the in-neighbor cosine-similarity score of the pair of nodes it is incident to. Although there is more information available that could be incorporated into the edge weights—a fraction of the Kaggle nodes have out-edges in addition to in-edges—we choose not to do so for simplicity. Further, one might imagine incorporating the intuition that higher-degree nodes are more likely to correspond to higher-degree nodes, but it is not clear if it is possible to encapsulate this insight in the form of edge weights.

---

[3]The "ground truth" mapping between obfuscated node IDs and real identities provided by Kaggle is not complete—it covers only 36,547 of the 37,692 nodes with outgoing edges, and 17,130 of the 17,594 nodes in the test set. The fractions reported in this paragraph are measured with respect to the available ground truth.

**Simulated annealing** is a state-space search "metaheuristic" that is ideal for approximate combinatorial optimization. The randomized search begins by exploring the state-space, placing relatively even probability on all directions, in order to avoid local optima. As the search proceeds, the distribution on directions begins to concentrate more tightly around the gradient direction, converging on greedy hill-climbing in the limit. A global parameter called *temperature*, which gradually decreases with time, controls the trade-off between exploration and hill-climbing. With a suitable "cooling schedule," local optima can be avoided in favor of the global optimum. For more see the tutorial survey [15].

Our goal in this section is to show that simulated annealing is a scalable approach to weighted graph matching, using a real-world dataset. The salient property of this dataset is that the difference between the graphs to be matched arises primarily due to the evolution of a social network with time. We are not claiming that simulated annealing is necessarily better suited than other approaches such as genetic algorithms, whether on this type of input or for weighted graph matching in general. We review the literature on graph matching in Section VI.

**Ground truth and dummy nodes.** Ideally, we would like to measure how close the output of our algorithm is to the global optimum, as well as how close the global optimum is to the "ground truth." However, we do not know the global optima, for obvious reasons. Therefore we directly measure the performance of the algorithm against the ground truth.

Another subtlety is that we actually want a partial mapping. We handle this by the typical approach of adding "dummy" nodes to both graphs [7].[4] Adding $k$ dummy nodes to $n$ regular nodes has the effect of finding a mapping of size $n - k$. This is because the weights of all edges incident on dummy nodes are 0, and therefore dummy nodes will never map to other dummy nodes (since that would result in a sub-optimal mapping that can be improved in 1 step).

If a mapping between two non-dummy nodes output by the algorithm is not an actual correspondence, we call it a false positive. If a correspondence between a pair of nodes is not output by the algorithm, it is a false negative. Ideally, we'd like the algorithm to behave as follows. Let $m$ be the number of pairs of corresponding nodes. Then the number of false positives should be $\max(n - m - k, 0)$ and the number of false negatives $\max(m + k - n, 0)$.

**Potential function.** Choosing a potential function is difficult—simply adding the differences between cosines gave poor results. For one, Figure 6 shows that a scaling factor is necessary. Moreover since a few false positives might be inevitable (with a sufficiently small number of

[4]This is analogous to *slack variables*.

**Input**:
$V_K, V_F$: ordered sets of Kaggle and Flickr nodes with the same index mapped to each other
$D_K, D_F$: sets of dummy nodes
$w_K, w_F$: weight functions over edges in Kaggle/Flickr
$i$: an index; $1 \leq i \leq n + k$
$\alpha$: parameter (set to 0.5)
$\beta$: parameter (set to 0.5)
**Output**: a distance measure

```
/* vectors of weights of edges between
node $i$ and non-dummy nodes */
```

$\sigma_K \leftarrow \langle w_K(i,j) \rangle_{1 \leq j \leq n+k,\ i \neq j,\ V_K[j] \notin D_K}$
$\sigma_F \leftarrow \langle w_F(i,j) \rangle_{1 \leq j \leq n+k,\ i \neq j,\ V_F[j] \notin D_F}$

$\bar{\sigma}_K \leftarrow \mathsf{mean}(\sigma_K)$
$\bar{\sigma}_F \leftarrow \mathsf{mean}(\sigma_F)$
**function** $\mathsf{PairDist}(x, y)$
   $r \leftarrow x > y\ ?\ x/y\ :\ y/x$
   **return** $(r - 1)^\alpha$
**return** $(\bar{\sigma}_K \bar{\sigma}_F)^{\frac{\beta}{2}} \sum_j \mathsf{PairDist}(\sigma_K[j]/\bar{\sigma}_K, \sigma_F[j]/\bar{\sigma}_F)$

**Algorithm 2:** Distance for simulated annealing. The potential function equals the distance summed over each of the indices. Since state transitions involve computing *differences* in the potential function, the summation is unnecessary in practice.

dummies), these might throw off the potential function due to wildly mismatched cosines. Finally, with the additive function, edges with higher cosine scores have a higher impact on the potential; we would like it to be more balanced.

The algorithm we used, listed as Algorithm 2, incorporates the above rationale. It compares a pair of mapped nodes; the potential function is the sum of the scores of the pairs of mapped nodes. We experimented with a few choices for $\alpha$ and $\beta$ and found that $\alpha = \beta = 0.5$ appears to yield the best results. The reason for this is not yet clear.

**States and cooling schedule.** The set of all bijective mappings between the two set of nodes (including the dummy nodes) forms the set of states for simulated annealing. Initially we start with a random bijection. The "neighbor states" of a bijection are derived by swapping the images of a pair of nodes.[5]

The acceptance probability of a state transition was chosen in a straightforward manner: $P(\Delta p, T) = e^{-\frac{\Delta p}{cT}}$ where $\Delta p$ is the change in potential function incurred by the transition, $T$ is the temperature, and $c$ is a constant. Thus, transitions to lower-potential states will always be taken, and other transitions will be taken with a nonzero probability.

$T$ was varied as $T = \frac{1}{t}$, where $t$ is the time or number of iterations. The "constant" $c$ is actually dependent on the number of nodes $n$: $c = c(n) = 20n$. These two choices

[5]We tried "clever" ways of picking neighbors where only pairs of nodes whose cosine similarity is greater than a threshold are considered. These choices performed worse than considering all possible pairs.

together constitute the "cooling schedule".

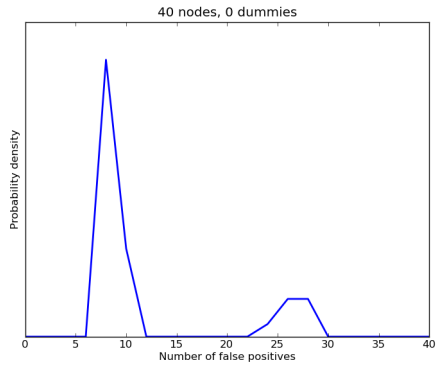

Fig. 8

ILLUSTRATION OF BIMODALITY. THIS BEHAVIOR IS TYPICAL

**Performance.** The results are summarized in Figures 9–11. All observations are medians of (at least) 30 trials. The median is a more meaningful statistic to report than the mean, because the distribution is often bimodal, with modes near 0 and $n$ as shown in Figure 8. This behavior is due to one of two events occurring: the algorithm gets stuck far away from the global optimum, or it finds the global optimum (or a point very close to it), which still results in some error because the global optimum differs from the ground truth by a small amount.

For $n = 20$, the algorithm matches the ideal performance (Figure 9). For $n = 40$, it is not quite ideal, but comes close, making no more than three errors (Figure 10). For $n = 80$, a different behavior is observed: when the number of dummies is too small, the algorithm is unable to find the global optimum in the median case.[6] With 8 or more dummies the algorithm performs well (Figure 11).

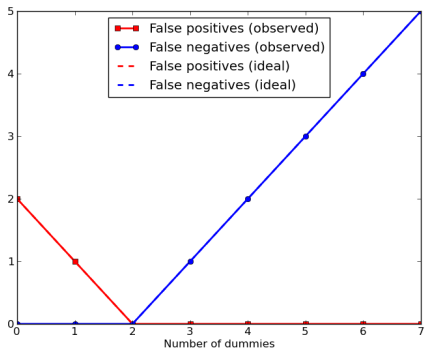

Fig. 9

SIMULATED ANNEALING PERFORMANCE: 20 NODES.

For the purposes of the propagation algorithm, the results

---

[6]With 0 dummies, the outputs are essentially no better than random permutations, although this is not shown in the graph.

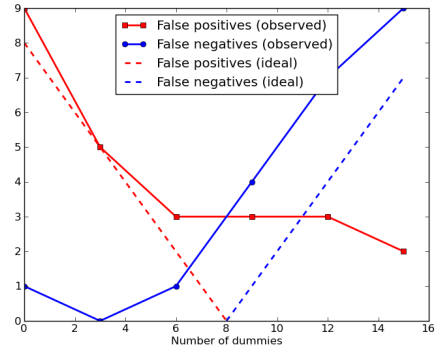

Fig. 10

SIMULATED ANNEALING PERFORMANCE: 40 NODES.

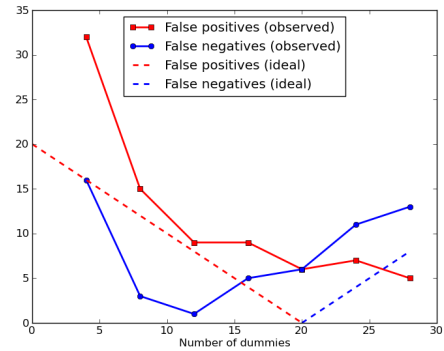

Fig. 11

SIMULATED ANNEALING PERFORMANCE: 80 NODES.

presented here are more than adequate, considering that it worked even with 10 seeds, and can tolerate a fraction of erroneous mappings.

**Parameters.** As the graphs show, the best results (measured by false positives + false negatives) are obtained when the selected number of dummies is (roughly) equal to the "correct" number of dummies, which is $n - m$, i.e. the number of nodes without a corresponding node in the other graph. Not knowing the optimal number of dummies is not a problem for our application, since we can try the propagation algorithm with different sets of seeds and as long as any one run results in an "epidemic", de-anonymization is successful.

Nevertheless, in other applications, being able to determine the right number of dummies automatically might be important. In combinatorial optimization problems, approaches that have fewer knobs that require tweaking are preferable. We describe two potential ways to accomplish this, both of which are possible future extensions:

- Pick the number of dummies conservatively and eliminate matches with high distance compared to the average.
- Run the algorithm in two phases. In the first phase, a high number of dummies is used, and in the second

phase a low number of dummies is used, but the non-dummy mapping obtained in the first phase is frozen.

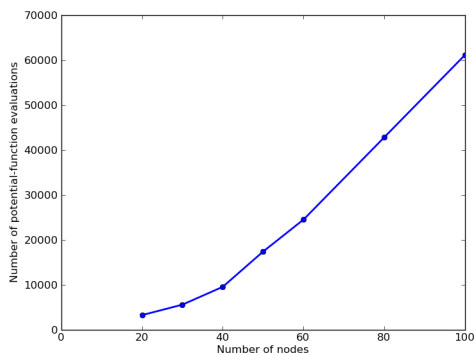

Fig. 12
SIMULATED ANNEALING RUNNING TIME.

**Running time.** Finally, we present some observations on running time. Figure 12 shows the number of evaluations of the potential function, which is the bottleneck step. We implemented a cache (with no expiry) of already-seen mappings and their potential function values; thus, what is shown on the $y$-axis is the number of cache misses. A further optimization computes only the difference in potential function values, by exploiting the fact that only one pair of nodes is swapped in one step, thus decreasing the complexity of potential function evaluation from quadratic to linear.

It is not clear how to interpret Figure 12. The curve appears to start off as a parabola and then become a straight line. We caution against generalizing too much from this figure, because in general the running time to get an equivalent error rate is likely to depend heavily on the nature of the data.

## V. LINK PREDICTION

### A. Use of voting to increase edge coverage

As stated earlier, de-anonymization covered 64.7% of the test cases in the test set. Recall that some nodes are not deterministically de-anonymized; multiple candidate matches are produced instead. We now describe how we leverage this in link prediction.

The idea behind voting is very simple. Let $v_K$ denote a Kaggle node, and $\mathcal{C}(v_K)$ denote $v_K$'s de-anonymization candidates, i.e., potential Flickr nodes that correspond to $v_K$. Given an edge $(a,b)$ in the test set, suppose $|\mathcal{C}(a)| \neq 0$ and $|\mathcal{C}(b)| \neq 0$. Then the edge $(a,b)$ has $|\mathcal{C}(a)| \cdot |\mathcal{C}(b)|$ possible de-anonymization candidates. If all candidates unanimously vote 0, i.e., none of these candidate edges exist in the Flickr graph, then we output the prediction 0. Conversely, if all candidates unanimously vote 1, we output the prediction 1.

To exploit the full power of voting, we performed the following additional steps to the de-anonymization algorithm: 1) Prune the de-anonymization output by confidence score. 2) Among the remaining nodes, run stage 2 of the propagation algorithm with the "sufficiently similar" criterion completely eliminated, thereby encouraging the algorithm to include even less likely candidates. As a result, some of the nodes have large numbers of candidates. This is not a drawback—since we adopt a unanimous voting strategy, having more candidates ensures that the predictions are more confident.

**Input**:

| $E_T$: | Test set |
| $E_F$: | Set of edges in the Flickr graph |
| $D$: | set of uniquely de-anonymized nodes |
| map: | a 1-1 de-anonymization mapping from Kaggle to Flickr nodes with domain $D$ |
| $\mathcal{C}(v)$: | set of de-anonymization candidates for Kaggle node $v$ |

**Output**: A prediction for each edge in $E_T$

**foreach** $(a,b) \in E_T$ **do**
    **if** $a \in D$ and $b \in D$ **then** /* DA */
        | **output** $(\mathsf{map}[a], \mathsf{map}[b]) \in E_F$?1 : 0
    **else if** $\mathcal{C}(a) \neq \emptyset$ and $\mathcal{C}(b) \neq \emptyset$ **then** /* Vote */
        pred ←
$$\begin{cases} 1 & \text{if } (\alpha,\beta) \in E_F \; \forall \alpha \in \mathcal{C}(a), \beta \in \mathcal{C}(b) \\ 0 & \text{if } (\alpha,\beta) \notin E_F \; \forall \alpha \in \mathcal{C}(a), \beta \in \mathcal{C}(b) \\ \textit{ML score} & \text{o.w.} \end{cases}$$
        **output** pred
    **else output** *ML score* ; /* ML */
**end**

**Algorithm 3:** Combining de-anonymization and machine learning.

Naturally, this gives an almost complete coverage of the nodes and edges, specifically 96.0% node coverage and 92.6% edge coverage. Of these 57.0% of edges had unique candidates; we applied voting to the other 35.6%. As shown in Table III, voting produced unanimous results for 18.7% of the 35.6% of edges.

Table I is a breakdown of the number of edges and non-edges found by the de-anonymization and voting strategy. One interesting observation is that the de-anonymization algorithm was able to determine more edges than non-edges, while the voting strategy uncovered only non-edges, 98.1% of which are true non-edges. This is not surprising, due to the following two reasons. First, the de-anonymization algorithm has a better coverage on higher-degree nodes. Second, we use a unanimous voting strategy to ensure high confidence. When a test edge has multiple de-anonymization candidates, it is highly likely that some candidates will vote no, as two nodes picked randomly from the graph are much more likely not to have an edge based on a simple calculation. Therefore, it is much more likely to uncover a non-edge through unanimous voting than to uncover an edge.

Algorithm 3 formally describes how we combine de-anonymization with machine learning predictions.

**Machine learning.** We implemented 25 features (and their variations) capturing neighborhood characteristics up to 4 hops away. These features include Adamic/Adar [19], Jaccard [19], localized random walks, node degrees, local clustering coefficients, whether the reverse edge exists, and so on. For a subset of the above features, we also implemented different variations with different parametrizations (e.g., hop count), and using in-degrees, out-degrees, or both.

TABLE I
RESULTS: breakdown of the number of edges and non-edges predicted by the de-anonymization and voting algorithms.

| Method | #(%) of edges predicted | #(%) of non-edges predicted |
|---|---|---|
| DA | 2868 (56.2%) | 2240 (43.8%) |
| Voting | 0 (0%) | 1677 (100%) |

We ran the Random Forest [8] classifier on top of the 25 selected features. We created a training and validation set using the "ground truth" obtained from de-anonymization. The training set contains $3,000 \sim 4,200$ examples, and the validation set contains roughly $1,000 \sim 2,000$ examples.

Table II depicts the performance of the machine learning algorithm, and selected features. Notably, among the features we tried, the best stand-alone feature is the localized random walk. Given a test edge $(a, b)$, we perform a random walk starting at node $a$ for a limited number of hops, and compute how likely it is to reach node $b$ in the process. To compute this probability, we implemented an approximate version of 3-4 rounds of matrix multiplication of the PageRank algorithm, starting at node $a$. The localized random walk feature achieves an AUC of $0.912$ and $0.924$ when the maximum number of hops is $3$ and $4$ respectively.

The Random Forest algorithm achieved an AUC of $0.935$ to $0.945$ on the validation set. On the entire Kaggle test set, it achieved an AUC of $0.953$; however, this number has to be taken with a grain of salt, as we used part of the test set for training (see Figure 13). The classifier performed worse on the ML set which was not covered by de-anonymization or voting, with an AUC of $0.881$. Therefore, an interesting observation is that the subset of test edges that are more difficult to de-anonymize also turns out to be the set relatively hard for link prediction (through ML). This is likely due to the fact that these nodes lack sufficient information necessary for de-anonymization or accurate link prediction.

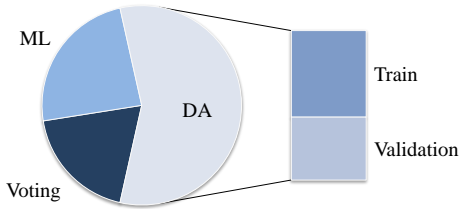

Fig. 13
BREAKDOWN OF THE KAGGLE TEST SET.

TABLE II
MACHINE LEARNING RESULTS.

| Method | AUC (test set) | AUC (ML set) |
|---|---|---|
| Adamic/Adar | 0.835 | 0.760 |
| Jaccard | 0.851 | 0.775 |
| Localized Random Walk (3 hops) | 0.912 | 0.834 |
| Localized Random Walk (4 hops) | 0.924 | 0.865 |
| Random Forest (25 features) | 0.953* | 0.881 |

*: Part of the test set was used to train based on the de-anonymization result.

TABLE III
RESULTS.

| Method | % Edges Covered | Accuracy | AUC |
|---|---|---|---|
| DA | 57.0%* | 0.987 | – |
| Voting | 18.7% | 0.981 | – |
| ML | 24.3% | – | 0.881** |
| **Overall** | 100% | | 0.981 |

\* Based on a pruned version of the de-anonymization result.
\*\* AUC on the ML set was 0.881; but AUC on the validation set was $0.935 \sim 0.945$. The ML set is a hard subset both for de-anonymization and machine learning since it is biased toward low-degree nodes.

## VI. RELATED WORK

**Link prediction.** Several prior works study the problem of social network link prediction through graph structural analysis [19], [23], [10]. Notably, our link prediction algorithm and those of other top teams leveraged several features proposed by Liben-Nowell and Kleinberg [19]. Unfortunately, the large size of the graph precluded the use of all features described therein when running on a commodity computer [11].

Other teams in the contest achieved impressive results using pure link prediction. In particular, Cukierski, Yang, and Hamner finished second, and their link prediction algorithm involved running Random Forests over $94$ selected features [11] and achieved an extraordinary AUC of $0.969$.

**Social network de-anonymization.** Backstrom et al. [6] showed how to de-anonymize specific nodes and edges in a social network by searching for patterns in subgraphs of small size (logarithmic; in practice, around $7$ nodes). Narayanan and Shmatikov studied de-anonymization of a social network on a large scale by using a different social network as auxiliary data [26].

**Inexact graph matching** is a well-studied problem that arises in diverse applications, primarily image recognition. Bengoetxea's thesis [7] surveys numerous approaches to solving it, including genetic algorithms, simulated annealing, expectation maximization, decision trees, and neural networks. It is possible that some of these techniques would extend to weighted graph matching.

Weighted graph matching has been studied in [29], [4]. These papers are concerned with finding the optimal solution rather than an approximation, and therefore their techniques only work for small numbers of nodes.

## VII. DISCUSSION

Our primary goal for participating in the Challenge was to raise attention to the ever-present possibility of de-anonymization in such contests. More broadly one might ask whether contests should discourage the use of outside information in general—trading off the benefits of leveraging outside information to genuinely improve the outcomes of data mining, and preventing cheating—and what technical or policy means could be used to limit such information.

In addition to undermining the aim of the contest, which was to advance the state-of-the-art in machine learning, de-anonymization also has privacy implications (although these did not arise in this contest). For these reasons, it is important to think about how to prevent de-anonymization in future contests. One factor that greatly affects both risks—in opposite directions—is whether the underlying data is already publicly available. If it is, then there is likely no privacy risk; however, it furnishes a ready source of high-quality data to game the contest. On the other hand, if the data is not public, then it can only be exploited by cross-referencing it through de-anonymization with another dataset of the same type (or with some of the same attributes) that *is* public.[7] While this presents a significant privacy risk, imperfect correlation between the two datasets makes it unclear whether there is a threat to the fidelity of the contest.

Focusing on the threat of de-anonymization to the contest, incorporating a restriction on the use of external data in the rules is a necessary step, but arguably insufficient, since less scrupulous contestants may attempt to apply de-anonymization anyway. We can see two possible ways to combat this threat. The first is the approach taken by the Netflix Prize—require teams qualifying for prizes to produce source code and human-readable descriptions of algorithms. The Netflix Prize stipulated that the verification process would "ensure that the provided algorithm description and source code could reasonably have generated the prediction sets submitted."

The loophole in this approach is the possibility of overfitting. While source-code verification would undoubtedly catch a contestant who achieved their results using de-anonymization alone, the more realistic threat is that of de-anonymization being used to bridge a small gap. In this scenario, a machine learning algorithm would be trained on the *test set*, the correct results having been obtained via de-anonymization. Since successful ML solutions are composites of numerous algorithms, and consequently have a huge number of parameters, it should be possible to conceal a significant amount of overfitting in this manner.

Another approach is to attempt to prevent the possibility of de-anonymization. This is a daunting task: a long line of research has demonstrated the feasibility of de-anonymization on various datasets [27], [21], [13], [17], [25], [26], [14], [1] and theoretical evidence exists for the impossibility of de-anonymization of high-dimensional data [3].

Several papers have claimed to provide k-anonymity, or variants thereof, for graphs [16], [20], [9], [30]. These have only been evaluated against graphs with a small number of nodes and small average degree. Further, they only consider the threat of adversaries with restricted types of auxiliary information, which in particular does not include *global* information such as another graph that is structurally related to the target graph. For these reasons we do not believe these techniques are effective ways to anonymize social networking graphs.

Finally, differential privacy offers a potential method to bypass the need for anonymization [12]. McSherry and Mironov show how the Netflix Prize contest could have been run by releasing a differentially private dataset (primarily, a sanitized item-item covariance matrix) instead of raw user data [22]. The drawbacks of this approach are that it would restrict the class of machine learning algorithms, and may require a shift from the "release-and-forget" model to an online privacy-preserving computation model, especially due to the difficulty of generating synthetic datasets that preserve differential privacy [28]. Nevertheless, differential privacy remains essentially the only mathematically sound methodology for provably resisting de-anonymization.

## VIII. Conclusions & Open Problems

We have described a de-anonymization attack on link prediction contests, which was used in the winning entry to the IJCNN 2011 Social Network Challenge run by Kaggle.com.

While the competition dataset was scrubbed of user identities, the revelation that the graph had been obtained by crawling Flickr proved sufficient for our attack. Using our own crawl of Flickr, we were able to successfully de-anonymize 64.7% of the test edge-set. And by training a Random Forest classifier on standard link prediction features of the training plus de-anonymized test sets, we were able to achieve a winning test AUC of 0.981.

We made several technical contributions in developing our attack. We introduced a new approach to seed-identification, which is applicable to weighted graph matching more generally. We also described how to combine the results of de-anonymization and link prediction: the link prediction training set is augmented by the partially de-anonymized test set, and probabilistic predictions from both processes are used together.

The success of our attack on the Social Network Challenge has important consequences for the future of such contests. We have argued that forbidding outside data sources and requiring source code submissions may not eliminate the problem. Ruling out de-anonymization while preserving meaningful utility remains an open question.

Identifying the source of anonymized social graph data is also an interesting research direction, although it is likely that global characteristics such as joint degree distribution would suffice to differentiate online social networks in a straightforward manner.

While previous work on social network link prediction mainly exploits the structure of the graph itself [19], [23], [10], our work suggests a promising new direction, namely, incorporating the vast amount of information publicly available on the web. Our de-anonymization approach may be regarded an extreme case; more generally, it may be possible to leverage statistical or aggregate information gleaned from the web to improve link prediction. It would also be interesting to investigate how to incorporate semantic information attached to nodes and edges into link prediction.

---

[7]For example, Narayanan and Shmatikov were able to re-identify anonymized records in the Netflix Prize dataset with user accounts on IMDb via their public movie reviews [25].

ACKNOWLEDGMENTS

We would like to thank Yong J Kil and Chris Li for their assistance and many helpful discussions. We are also very grateful to the competition organizers Dirk Nachbar, Anthony Goldbloom, Jeremy Howard, and the collegial atmosphere shared by our fellow Challenge participants. Elaine Shi would like to thank Srinath Sridhar and Chengwen Chris Wang for their helpful consultations.

REFERENCES

[1] A. Abbasi and H. Chen. Writeprints: A stylometric approach to identity-level identification and similarity detection in cyberspace. *ACM Trans. Inf. Sys.*, 26:7:1–7:29, April 2008.
[2] A. M. Abdulkader. *Parallel Algorithms for Labelled Graph Matching*. PhD thesis, Colorado School of Mines, 1998.
[3] C. C. Aggarwal. On k-anonymity and the curse of dimensionality. In *In VLDB*, pages 901–909, 2005.
[4] H. Almohamad and S. Duffuaa. A linear programming approach for the weighted graph matching problem. *IEEE Trans. Pattern Anal. Mach. Intell.*, 1993.
[5] S. F. Altschul, W. Gish, W. Miller, E. W. Myers, and D. J. Lipman. *J. Molecular Biol.*, 1990.
[6] L. Backstrom, C. Dwork, and J. Kleinberg. Wherefore art thou r3579x?: anonymized social networks, hidden patterns, and structural steganography. In *Proc. World Wide Web*, pages 181–190, 2007.
[7] E. Bengoetxea. *Inexact Graph Matching Using Estimation of Distribution Algorithms*. PhD thesis, Ecole Nationale Supérieure des Télécommunications, Paris, France, Dec 2002.
[8] L. Breiman. Random forests. *Machine Learning*, 45(1):5–32, 2001.
[9] A. Campan and T. M. Truta. A clustering approach for data and structural anonymity in social networks. In *In Privacy, Security, and Trust in KDD Workshop (PinKDD)*, 2008.
[10] D. Corlette and F. M. Shipman, III. Link prediction applied to an open large-scale online social network. In *Proceedings of the 21st ACM conference on Hypertext and hypermedia*, HT '10, 2010.
[11] W. Cukierski, B. Yang, and B. Hamner. How I did it: Will Cukierski on finishing second in the IJCNN Social Network Challenge. http://www.kaggle.com/blog/?p=728, 2010.
[12] C. Dwork. Differential privacy: a survey of results. In *TAMC*, 2008.
[13] D. Frankowski, D. Cosley, S. Sen, L. Terveen, and J. Riedl. You are what you say: privacy risks of public mentions. In *SIGIR*, 2006.
[14] P. Golle and K. Partridge. On the anonymity of home/work location pairs. In *Pervasive*, pages 390–397, 2009.
[15] B. Hajek. A tutorial survey of theory and applications of simulated annealing. In *IEEE Conference on Decision and Control*, 1985.
[16] M. Hay, G. Miklau, D. Jensen, D. Towsley, and P. Weis. Resisting structural re-identification in anonymized social networks. *Proc. VLDB Endow.*, 1:102–114, August 2008.
[17] N. Homer, S. Szelinger, M. Redman, D. Duggan, W. Tembe, J. Muehling, J. V. Pearson, D. A. Stephan, S. F. Nelson, and D. W. Craig. Resolving individuals contributing trace amounts of DNA to highly complex mixtures using high-density SNP genotyping microarrays. *PLoS Genet*, 4(8):e1000167, 08 2008.
[18] Kaggle.com. IJCNN social network challenge. http://www.kaggle.com/socialNetwork, 2011.
[19] D. Liben-Nowell and J. Kleinberg. The link prediction problem for social networks. In *Proc. CIKM '03*, pages 556–559, 2003.
[20] K. Liu and E. Terzi. Towards identity anonymization on graphs. In *SIGMOD Conference*, pages 93–106, 2008.
[21] B. Malin and L. Sweeney. How (not) to protect genomic data privacy in a distributed network: using trail re-identification to evaluate and design anonymity protection systems. *J. Biomedical Informatics*, 37:179–192, June 2004.
[22] F. McSherry and I. Mironov. Differentially private recommender systems: Building privacy into the netflix prize contenders. In *KDD'09*, pages 627–636, 2009.
[23] T. Murata and S. Moriyasu. Link prediction based on structural properties of online social networks. *New Generation Comput.*, 26(3):245–257, 2008.
[24] D. Nachbar. Personal communication, 2011.
[25] A. Narayanan and V. Shmatikov. Robust de-anonymization of large sparse datasets. *IEEE Symp. Security and Privacy*, 0:111–125, 2008.
[26] A. Narayanan and V. Shmatikov. De-anonymizing social networks. *IEEE Symp. Security and Privacy*, 0:173–187, 2009.
[27] L. Sweeney. k-anonymity: A model for protecting privacy. *Int. J. Uncertainty, Fuzziness and Knowledge-Based Sys.*, 2002.
[28] J. Ullman and S. P. Vadhan. Pcps and the hardness of generating synthetic data. *Electronic Colloquium on Computational Complexity (ECCC)*, 17:17, 2010.
[29] S. Umeyama. An eigendecomposition approach to weighted graph matching problems. *IEEE Trans. Pattern Anal. Mach. Intell.*, 1988.
[30] B. Zhou and J. Pei. Preserving privacy in social networks against neighborhood attacks. In *ICDE*, pages 506–515, 2008.